\documentclass[epj]{svjour}
\usepackage[dvips]{graphicx}
\begin{document}
\title{Imprints of energy limitation in
transverse momentum distributions of jets}
\author{Maciej Rybczy\' nski 
\thanks{\emph{e-mail: maciej.rybczynski@ujk.edu.pl}}
\and Zbigniew W\l odarczyk
\thanks{\emph{e-mail: zbigniew.wlodarczyk@ujk.edu.pl}}
}                     
\institute{Institute of Physics,
                Jan Kochanowski University,  \'Swi\c{e}tokrzyska 15,
                25-406 Kielce, Poland}
\date{Received: date / Revised version: date}
%
\abstract{ Using a Tsallis nonextensive approach, we analyse distributions 
of transverse spectra of jets. We discuss the possible influence of
energy conservation laws on these distributions. Transverse spectra of jets
exhibit a power-law behavior of $1/p_T^n$ with the power indices $n$ similar to
those for transverse spectra of hadrons.
} 
\authorrunning{M. Rybczy\' nski and Z. W\l odarczyk}
\titlerunning{Transverse momentum of jets ...}

\maketitle


\section{Introduction}
\label{sec:Introduction}

For some time now it is known that transverse momentum spectra of different
kinds measured in multiparticle production processes can be described by 
a quasi power-law formula~\footnote{Throughout the whole article a natural units 
(Boltzmann constant and speed of light) are used by setting its values to unity.},
\begin{equation}
h(p_T) = C \left( 1 + \frac{p_T}{nT} \right)^{-n}, 
\label{eq:hagedorn}
\end{equation}
which for large values of transverse momenta, $p_T \gg nT$, becomes scale free
(independent of $T$) power distribution, $ 1/p_T^n$. This was first proposed in~\cite{CM1,CM2} as the simplest formula extrapolating exponential behavior observed for low $p_T$ to power behavior at large $p_T$. At present it is known as the QCD-inspired {\it Hagedorn formula} \cite{CM3,CM4}. Distribution~(\ref{eq:hagedorn}), nowadays, has been successfully used for a description of multiparticle production processes in a wide range of incident energy
(from a few GeV up to a few of TeV) and in broad range of transverse momenta (see, for example, reviews~\cite{WWrev1,WWrev2,WWrev3}). In particular, it turned out that it successfully describes a transverse momenta of charged particles measured by the LHC experiments, the flux of which changes by over 14 orders of magnitude~\cite{qQCD1,qQCD2,qWWCT2,qWWCT}.

In many branches of physics Eq.~(\ref{eq:hagedorn}), with $n$ replaced by $n=1/(q-1)$, it is widely known as the {\it Tsallis formula}~\cite{Tsallis1,Tsallis2,Tsallis3}. In this form, Eq.~(\ref{eq:hagedorn}) is usually supposed to represent a nonextensive generalization of the Boltzmann-Gibbs exponential distribution, $exp(-p_T/T)$, used in a statistical description of multiparticle production processes. Such an approach is known as nonextensive statistics~\cite{Tsallis1,Tsallis2,Tsallis3} in which the nonextensivity parameter $q$ summarily describes all features causing a departure from the usual Boltzmann-Gibbs statistics. In particular it can be shown that the nonextensivity parameter $q$ is directly related to the fluctuations of the parameter $T$ identified with the "temperature" of the hadronizing fireball~\cite{WWq1,WWq2}.

It was shown there that such a situation can occur when the heat bath is not homogeneous and must be described by a local temperature, $T'$, fluctuating from point to point around some equilibrium value, $T$. Assuming some simple diffusion picture as being responsible for equalization of this temperature~\cite{WWrev1,WWq1,WWq2} one obtains the evolution of $T'$ in the form of a Langevin stochastic equation with the distribution of $1/T'$, $g(1/T')$, emerging as a solution of the corresponding Fokker-Planck equation. It turns out that in this case $g(1/T')$ takes the form 
of a gamma distribution,
\begin{eqnarray}
g(1/T') &=& \frac{nT}{\Gamma\left(n\right)}\left(
\frac{nT}{T'}\right)^{n-1} 
\exp\left( - \frac{nT}{T'}\right).
\label{eq:gamma}
\end{eqnarray}
After convolution of the usual Boltzmann-Gibbs exponential factor $\exp (-p_T/T')$ with Eq.~(\ref{eq:gamma}) 
one immediately obtains a Tsallis distribution, Eq.~(\ref{eq:hagedorn}), with
\begin{equation}
n= \frac{<T'>^2}{Var(T')}, \label{eq:q}
\end{equation}
directly connected to the variance of $T'$. This idea was further developed in~\cite{BJ1,BJ2,B} (where problems connected with the notion of temperature in such cases were addressed). The above makes a basis for the so-called {\it superstatistics}~\cite{SuperS1,SuperS2}.

Applications of Tsallis distributions to multiparticle production processes are now numerous. To those quoted previously in~\cite{WWrev1} one should add some new results from~\cite{WWprc,BPU_epja,WWrev2} and also presented in~\cite{B}. The most recent applications of this approach come from the STAR and PHENIX Collaborations at RHIC~\cite{STAR,PHENIX} and from CMS~\cite{CMS1,CMS2}, ALICE~\cite{ALICE1,ALICE2} and ATLAS~\cite{ATLAS} Collaborations at LHC (see also a recent compilations~\cite{qcompilation1,qcompilation2}).

It is now empirically well-documented that transverse spectra of both hadrons and jets exhibit a power-law behavior of $1/p_T^n$ at hight $p_T$. This observation (usually interpreted in terms of non-extensive Tsallis statistics) meets difficulty. Whereas in~\cite{qQCD2,qWWCT2} it has been advocated that the power indices $n$ for hadrons are systematically greater than those for jets, in~\cite{WWPLB} the values of corresponding power indices were found to be
similar, strongly indicating the existence of a common mechanism behind all these processes.

In Section~\ref{section:Conditional} the influence of constrains, forcing the use of conditional probabilities for Tsallis distribution (and the case of Boltzmann statistics in Appendix) is discussed. In Section~\ref{sec:Imprints} we report the results concerning the consequences of the use of conditional Tsallis distribution for description of transverse spectra of jets. Section~\ref{sec:Summary} is our summary and discussion of results.


\section{Conditional probability - an influence of conservation laws}
\label{section:Conditional}

\begin{figure}[h]
\begin{center}
\resizebox{0.48\textwidth}{!}{
  \includegraphics{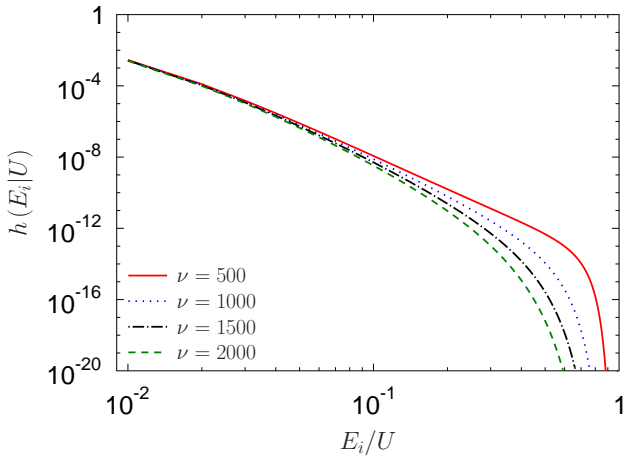}
  }
\caption{(Color online) Conditional probability distribution, $h\left(E_i|U \right)$, as a function of of $ E_i/U$ for Tsallis statistics ($n=7$ and $T=1 $ GeV).
 } \label{Fig_conditional_1}
\end{center}
\end{figure}
\begin{figure}[h]
\begin{center}
\resizebox{0.48\textwidth}{!}{
  \includegraphics{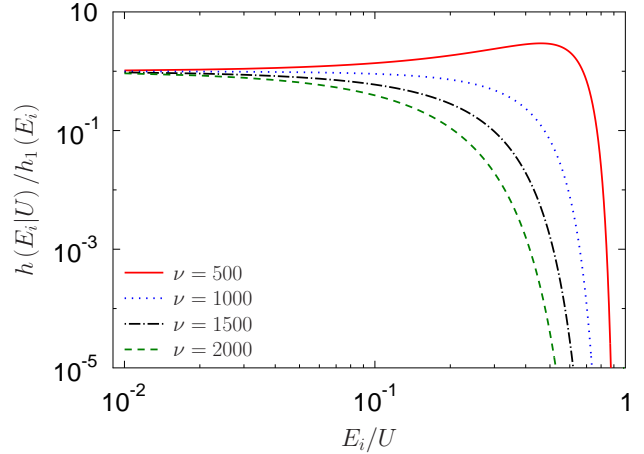}
  }
\caption{(Color online) Ratio of the conditional distribution function $h(E_i|U)$ and a single particle distribution $h_{1}(E_i)$ as a function of $E_i/U$, for Tsallis statistics ($n=7$ and $T = 1$ GeV).
 } \label{Fig_conditional_2}
\end{center}
\end{figure}

Let $\{ E_{1,\dots, \nu}\}$ be a set of $\nu$ independent identically distributed random variables described by some parameter $T$. For independent energies, $\{E_{i=1,\dots, \nu}\}$, each distributed according to the simple Boltzmann distribution:
\begin{equation}
g_1\left(E_i\right) = \frac{1}{T} \exp\left( - \frac{E_i}{T}\right), 
\label{eq:Boltzmann}
\end{equation}
the sum
$U = \sum^{\nu}_{i=1} E_i$
is then distributed according to the distribution,
$g_{\nu}(U) = 
[T \Gamma{(\nu )}]^{-1}(U/T)^{\nu-1} \exp(-U/T)$.

If the available energy $U$ is limited, for example if $U = \sum^{\nu }_{i=1} E_i =  const$, then we have the following conditional probability for the single particle distribution, $g\left(E_i\right)$:
\begin{eqnarray}
g\left( E_i|U  \right) &=& \frac{g_1\left(E_i\right)
g_{\nu -1}\left(U - E_i\right)}{g_{\nu}(U)} =\nonumber\\
&=& \frac{(\nu - 1)}{U}\left(1 - 
\frac{E_i}{U} \right)^{\nu  - 2}. 
\label{eq:condprob1}
\end{eqnarray}
In the above consideration the value of the $T$ parameter {\it does not fluctuate}. More details is given in 
Appendix and review~\cite{WWAPP}.

Now let us consider a situation in which the parameter $T$ in the joint probability distribution 
\begin{equation}
g\left( \{E_{1,\dots,\nu}\}\right) = \prod^{\nu}_{i=1}g_1\left( E_i\right)
\label{eq:joint}
\end{equation}
fluctuates according to a gamma distribution, Eq.~(\ref{eq:gamma}). In this case we have a single particle Tsallis distribution
\begin{equation}
h_i\left(E_i\right) = \frac{n-1}{n T} \left( 1 + \frac{E_i}{n T}\right)^{-n} 
\label{eq:hTsallis}
\end{equation}
and a distribution of  $U=\sum^{\nu}_{i=1} E_i$  is given by (cf.~\cite{WWcov}):
\begin{eqnarray}
h_{\nu}(U) &=& \frac{\Gamma\left( \nu + n-1\right)}
{U \Gamma(\nu) \Gamma\left( n-1\right)} 
\left( \frac{U}{T}\right)^{\nu}\left( 1 + 
\frac{U}{n T} \right)^{1 - \nu -n}.
\label{eq:hNE}
\end{eqnarray}
If  the energy is limited, i.e., if $U = \sum^{\nu}_{i=1} E_i  = const$, we have the following conditional probability:
\begin{eqnarray}
h\left( E_i|U\right) &=& \frac{ h_1\left( E_i\right)h_{\nu -1}\left(U
- E_i\right)}{h_{\nu}(U)} = \nonumber\\
&=& \frac{(\nu - 1)(n-1)}{(n-2 + \nu)} \frac{(nT+U)}{nTU}
\left( \frac{U - E_i}{U}\right)^{\nu -1} \times \nonumber\\
& \times & \left( 1 +\frac{E_i}{n T}\right)^{-n}\left( 1 -\frac{E_i}{n T+U}\right)^{2 - \nu -n}.
\label{eq:fEiE}
\end{eqnarray}

For $n \rightarrow \infty$ Eq. (\ref{eq:fEiE}) reduces to Eq.~(\ref{eq:condprob1}). On the other hand, for large energy ($U\rightarrow \infty$) and large number of degrees of freedom ($\nu \rightarrow\infty$), the conditional probability distribution (\ref{eq:fEiE}) reduces to the single particle distribution given by Eq.~(\ref{eq:hTsallis}). 

For $E_i \ll U$ the conditional probability (\ref{eq:fEiE}) can be rewritten as 
\begin{eqnarray}
h\left( E_i|U\right) &\simeq& \frac{\left(\nu +2\right)(n-1)}
{n T(n-2 + \nu)} \left(1 +\frac{E_i}{n T}\right)^{-n} 
\label{eq:eismallaE}
\end{eqnarray}
which, when additionally $\nu \gg 1$,  reduces to Eq.~(\ref{eq:hTsallis}).

The results presented here are summarized in Figs.~\ref{Fig_conditional_1} and~\ref{Fig_conditional_2} 
which show how large are differences between the {\it conditional} Tsallis distribution 
$h(E_i|U)$ and the {\it usual} $h_1(E_i)$.


\section{Imprints of constrains in transverse spectra} 
\label{sec:Imprints}

\begin{figure}[h]
\begin{center}
\resizebox{0.48\textwidth}{!}{
  \includegraphics{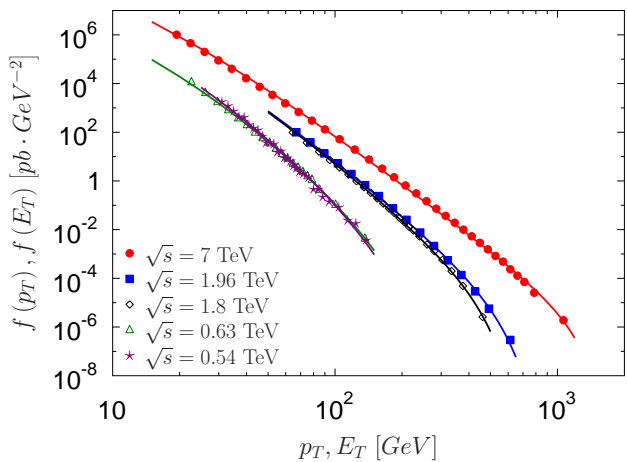}
  }
\caption{(Color online) Transverse spectra for jets ($p_T$ and $E_T$ distribution
are shown by full and open symbols, respectively) fitted by the
conditional Tsallis distribution given by Eq.~(\ref{eq:fEiE}).
 } 
\label{Fig_spectra}
\end{center}
\end{figure}

\begin{figure}[h]
\begin{center}
\resizebox{0.48\textwidth}{!}{
  \includegraphics{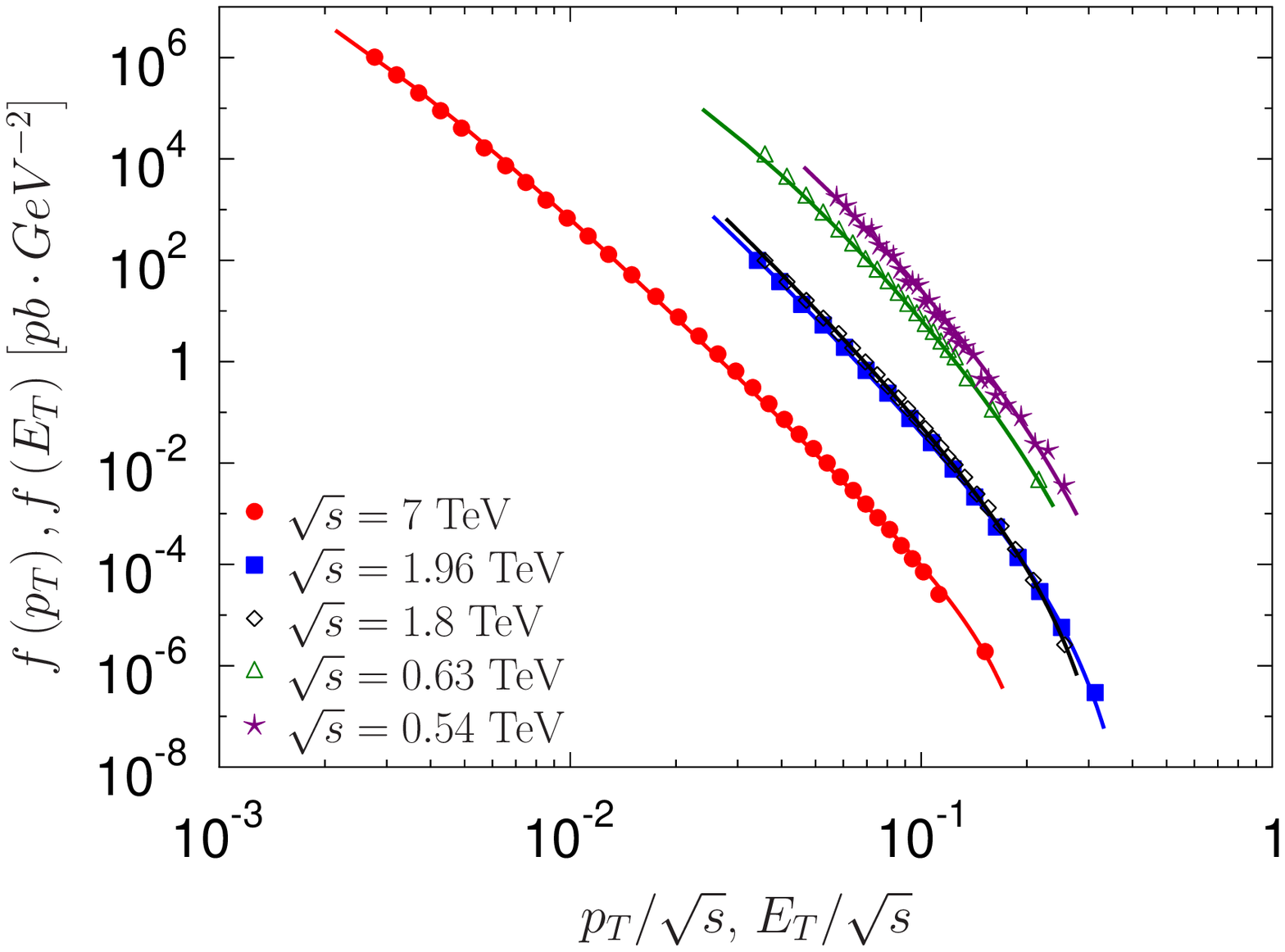}
  }
\caption{(Color online) Jet differential yields as a function of $p_T/\sqrt{s}$ or $E_T/\sqrt{s}$ ($p_T$ and $E_T$ distributions are shown by full and open symbols, respectively) fitted by the conditional Tsallis distribution given by Eq.~(\ref{eq:fEiE}). 
 } 
\label{Fig_spectra_sqrts}
\end{center}
\end{figure}

\begin{figure}[h]
\begin{center}
\resizebox{0.48\textwidth}{!}{
  \includegraphics{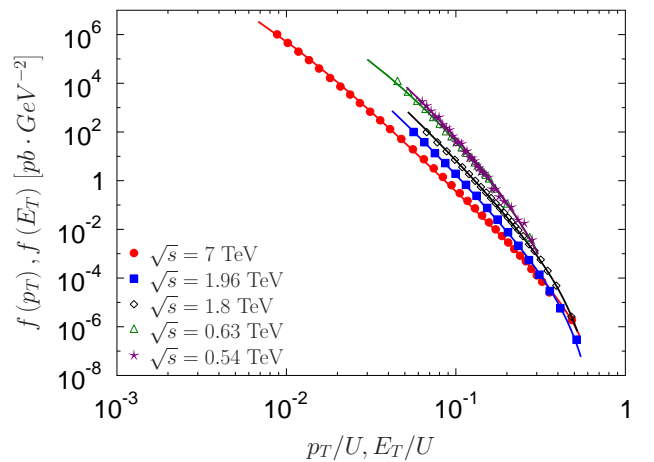}
  }
\caption{(Color online) Jet differential yields as a function of $p_T/U$ or $E_T/U$ ($p_T$ and $E_T$ distribution
are shown by full and open symbols, respectively) fitted by the conditional Tsallis distribution given by Eq.~(\ref{eq:fEiE}). 
 } 
\label{Fig_spectra_u}
\end{center}
\end{figure}

Now we concentrate on the experimental distributions of transverse momenta 
\begin{equation}
f(p_T) =\frac{d^2 \sigma}{p_Tdp_Td\eta}  
\label{eq:spectrum}
\end{equation}
of jets observed at midrapidity and in a narrow jet cone defined by $R=\sqrt{\Delta\eta^2+\Delta\phi^2}$ (where
 $\Delta\phi$ and $\Delta\eta$ are, respectively, the azimuthal angle and the pseudorapidity of hadrons relative
to that of the jet). Data on $p+\bar{p}$ and $p+p$ interactions, covering a wide energy range from $0.54$~TeV 
up to $7$~TeV, obtained by CMS~\cite{CMS:2011ab}, CDF II~\cite{Aaltonen:2008eq}, D0~\cite{Abbott:2000kp} and UA2~\cite{Bagnaia:1984pn} experiments was used for the analysis (cf. Table~\ref{Tab:desc} for more details). 
Because data, which we analyze, are presented at midrapidity (i.e., for $\eta \sim  0$) and for large transverse momenta, $p_T  \gg \mu$, the energy $E$ of a produced jets is roughly equal to its transverse momentum $p_T$ (and transverse energy  $E_T$), which we shall use in what follows.

Transverse spectra for jets $f(p_T)$, fitted by the conditional Tsallis distribution $h(p_T|U)$ given by Eq.~(\ref{eq:fEiE}), are shown in Fig.~\ref{Fig_spectra}. Parameters of the Tsallis distribution used to fit the data are shown in Table~\ref{Tab:params}. An available energy per degree of freedom
\begin{equation}
\frac{U}{\nu}\cong 0.22\cdot \ln \left( \frac{\sqrt{s}}{33}  \right)
\label{um}
\end{equation}
increase logarithmically with energy $\sqrt{s}$, while $\nu / \sqrt{s}\cong 12.7-2.73 \ ln \sqrt{s} +0.15 \ (ln \sqrt{s})^2 $ and $U/ \sqrt{s} =2.35-0.23 \ ln \sqrt{s}$.
The slope parameters decrease with energy as
\begin{equation}
n=12.25 \left( \sqrt{s} \right)^{-0.064}
\label{n}
\end{equation}
and numerically are comparable with $n(s)=1/(q(s)-1)$ for $q(s)$ evaluated for transverse spectra of charged 
particles (for charge particles $n=14.78 (\sqrt{s})^{-0.09}$)~\cite{Wibig}.

Of course many other parametrisations, than this given by Eq.~(\ref{n}), is possible in our limited interval of energy (0.5 TeV - 7 TeV). For example $n=4+10(\sqrt{s})^{-0.137}$ for jets and $n=4+15.6(\sqrt{s})^{-0.2}$ for charged particles. However all parametrisations lead to conclusions that $n_{jet}=n_{char}$ at TeV energies (1.2 TeV - 1.4 TeV).

The flux at given $p_T$ (or $E_T$) increase with interaction energy $\sqrt{s}$ as
\begin{equation}
f(p_T=100 \ GeV)=5.5\cdot 10^{-6} (\sqrt{s})^{1.85}-0.6 .
\label{flux}
\end{equation}
However a comparison of fluxes at transverse momenta close to the interaction energy (cf. Fig.~\ref{Fig_spectra_sqrts} show decrease with energy
\begin{equation}
f(p_T/ \sqrt{s}=0.1 )=10^{14} (\sqrt{s})^{-4.7}.
\label{flux_s}
\end{equation}
With increasing interaction energy, jets are produced with lower relative transverse momenta and paradoxically for lower $\sqrt{s}$ we are closer to the kinematical limits. For transverse momenta close to available energy $U$, presented in Fig.~\ref{Fig_spectra_u}, fluxes of jets show rapid decrease with $p_T/U \rightarrow 1$.

\begin{table}
\caption{ Jet production in $p+\bar{p}$ and $p+p$ interactions.}
\label{Tab:desc}
\centering
\begin{tabular}{||c|c|c|c|c|c||}\hline
Exp. & Reaction & Variable& $\sqrt{s}$~[TeV] & $R$ & $|\eta|$\\ \hline
CMS & $p+p$ & $p_T$ & $7.0$ & $0.5$ & $<0.5$\\ \hline
CDF II &$p+\bar{p}$ & $p_T$ & $1.96$ & $0.7$ & $<0.1$ \\ \hline
D0 & $p+\bar{p}$  & $E_T$ & $1.8$ & $0.7$ & $<0.5$\\ \hline
D0 & $p+\bar{p}$ & $E_T$ & $0.63$ & $0.7$ & $<0.5$\\ \hline
UA2 & $p+\bar{p}$ & $p_T$ & $0.54$ & - & $<0.85$ \\ \hline
\end{tabular}
\end{table} 

\begin{table}
\caption{Parameters of the Tsallis distribution used to fit the data.}
\label{Tab:params}
\centering
\begin{tabular}{||c|c|c|c|c|c||}\hline
Exp. & $\sqrt{s}$~[TeV] & $n$ & $m$ & $T$~[GeV] & $U$~[GeV]\\ \hline
CMS & $7.0$ & $6.95$ & $1900$ & $1.08$ & $2200$\\ \hline
CDF II & $1.96$ & $7.55$ & $1340$ & $1.08$ & $1190$\\ \hline
D0 & $1.8$ & $7.7$ & $1050$ & $1.08$ & $960$\\ \hline
D0 & $0.63$ & $8.0$ & $800$ & $1.08$ & $500$\\ \hline
UA2 & $0.54$ & $8.3$ & $800$ & $1.08$ & $490$\\ \hline
\end{tabular}
\end{table} 


\section{Summary}
\label{sec:Summary}

Transverse momentum spectra of jets, spanning over $12$ orders of magnitude in the observed cross sections,
can be successfully described by the conditional Tsallis distribution. Those spectra exhibit a power-law bahavior of $1/p_T^n$ for which $n\approx 7-8$ ($n \approx 8$ at 800 GeV and decrease with energy, $n=8 (\sqrt{s}/800)^{-0.064}$.  
Observed deviation from the power-law at tail of distribution is caused by kinematical constrains.
Fluxes of jets for $p_T/\sqrt{s}=const$ are higher for lower incident energies, and are comparable for
high $p_T/U$ where fluxes have been lowering seemingly.

\begin{figure}[h]
\begin{center}
\resizebox{0.48\textwidth}{!}{
  \includegraphics{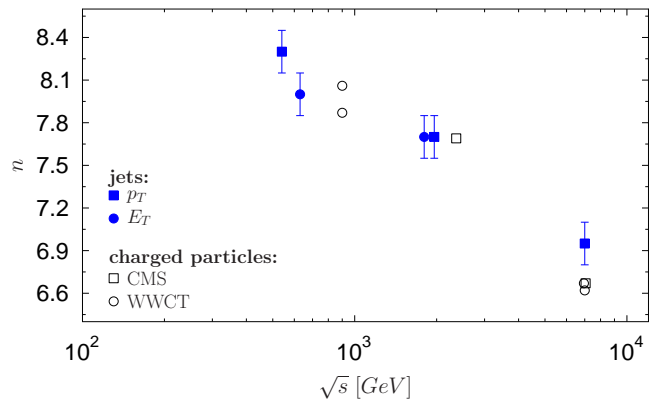}
  }
\caption{(Color online) Slope parameters $n$ extracted from the fits to the transverse spectra of jets (full symbols) in comparison with slope parameters for distributions of charged particles (open symbols: CMS~\cite{CMS1,CMS2} and WWCT~\cite{qWWCT}).
 } 
\label{Fig_n}
\end{center}
\end{figure}

Comparison of power indices $n$ for jets with those for charged particles produced in minimum bias collisions is 
shown in Fig.~\ref{Fig_n}. The values of the corresponding power indices are similar, indicating the existence 
of a common mechanism behind all these processes~\cite{WWPLB}. The suggestion that power indices for jets are 
almost 2 times smaller than those for hadrons~\cite{qQCD2,qWWCT2}, comes mainly from a different parametrisation of hadrons and jets spectra (not a simple Tsallis distribution) and can not be accepted by looking directly at the 
corresponding distributions.

\renewcommand{\theequation}{A-\arabic{equation}}
\setcounter{equation}{0}  
\section*{Appendix A: Conditional Boltzmann distribution}  
\label{sec:Appendix}

For independent energies, $\{E_{i=1,\dots,\nu}\}$, each distributed according to the simple Boltzmann distribution, Eq.~(\ref{eq:Boltzmann}) the sum
\begin{equation}
U = \sum^{\nu}_{i=1} E_i \label{eq:SumEi}
\end{equation}
is distributed accordingly to the following gamma distribution,
\begin{equation}
g_{\nu}(U) = \frac{1}{T(\nu - 1)!}\left(
\frac{U}{T}\right)^{\nu-1}\exp\left(
-\frac{U}{T}\right). \label{eq:gammaE}
\end{equation}
If the available energy $U$ is limited, for example if $U = const$, then we have the following
{\it conditional probability} for the single particle distribution (cf. Eq.~(\ref{eq:condprob1})):
\begin{eqnarray}
g\left( E_i|U \right) 
&=& \frac{(\nu - 1)}{U} \left(1 - 
\frac{E_i}{U} \right)^{\nu - 2}. \label{eq:condprob}
\end{eqnarray}
This is nothing else than the well-known Tsallis distribution
\begin{equation}
g\left(E_i | U \right) = \frac{2 - q}{T} \left[ 1 -
(1 - q)\frac{E_i}{T}\right]^{\frac{1}{1 - q}}
\label{eq:condTsallis}
\end{equation}
with nonextensivity parameter
\begin{equation}
q = \frac{\nu - 3}{\nu - 2} 
\label{eq:parcond}
\end{equation}
which is always less than unity and $T=U/(\nu -1)$.

Actually, such distributions emerge directly from the calculus of probability for situation known as {\it
induced partition}~\cite{IndPart}. In short: $\nu-1$ randomly chosen independent points $\left\{ U_{1},\dots, U_{\nu-1}\right\}$ breaks segment $(0,U)$ into $\nu$ parts, length of which is distributed according to Eq.~(\ref{eq:condprob}). The length of the $k^{th}$ such part corresponds to the value of energy 
$E_k = U_{k+1}-U_k$ (for ordered $U_k$). One could think of some analogy in physics to the case of random breaks of string in $\nu-1$ points in the energy space. 

We end this part by a reminder of how Tsallis distribution with $q<1$ arises from {\it statistical physics considerations}. Consider an isolated system with energy $U=const$ and with $\nu$ degrees of freedom. Choose a single degree of freedom with energy $E$ (i.e. the remaining or reservoir energy is $E_r = U - E$). If this degree of freedom is in a single, well defined state, then the number of states in the whole system equals $\Omega(U-E)$. The  probability that the energy of the chosen degree of freedom equals to $E$ is then $P(E) \propto \Omega(U-E)$. Expanding (slowly varying) $\ln \Omega(E)$ around $U$,
\begin{equation}
\ln \Omega(U-E) = \sum_{k=0}^{\infty} \frac{1}{k!}
\frac{\partial^{(k)} \ln \Omega}{\partial E_r^{(k)}}, \quad 
\label{eq:deriv}
\end{equation}
and (because $E\ll U$) keeping only the two first terms one gets $\ln P(E) \propto \ln \Omega(E) \propto - \beta E$, i.e., $P(E)$ is a Boltzmann distribution given by Eq.~(\ref{eq:Boltzmann}) with
\begin{equation}
\frac{1}{ T} =\beta \stackrel{def}{=}
\frac{\partial \ln \Omega\left(E_r\right)}{\partial E_r}.
\label{eq:beta}
\end{equation}

On the other hand, because one usually expects that $\Omega\left(E_r\right) \propto \left(E_r/\nu\right)^{\alpha_1 \nu - \alpha_2}$ (where $\alpha_{1,2}$ are of the order of unity and we put $\alpha_1 = 1$ and, to account for diminishing the number of states in the reservoir by one, $\alpha_2 = 2$) \cite{Reif,Feynman}, one can write
\begin{equation}
\frac{\partial^k \beta}{\partial E_r^k} \propto (-1)^k k!
\frac{\nu - 2}{E^{k+1}_r} = (-1)^k k! \frac{\beta^{k-1}}{(\nu -
2)^k} \label{eq:FR}
\end{equation}
and write the full series for probability of choosing energy $E$:
\begin{eqnarray}
P(E)&\propto& \frac{\Omega(U-E)}{\Omega(U)} = \exp\left[
\sum_{k=0}^{\infty}\frac{(-1)^k}{k+1}\frac{(- \beta E)^{k+1}}{(\nu - 2)^k}
\right]=\nonumber\\
&=& C\left(1 - \frac{1}{\nu - 2}\beta E\right)^{(\nu - 2)}
=\nonumber\\
&=& \beta(2-q)[1 - (1-q)\beta E]^{\frac{1}{1-q}};
\label{eq:statres}
\end{eqnarray}
(where we have used the equality $\ln(1+x) = \sum_{k=0}^{\infty}(-1)^k[x^{k+1}/(k+1)]$). This result, with 
$q = (\nu-3)/(\nu -2) \leq 1$, coincides with the results from conditional probability and the induced partition. 


\end{document}